\documentclass[pdflatex,sn-nature,iicol]{sn-jnl}

\usepackage{graphicx}%
\usepackage{multirow}%
\usepackage{amsmath,amssymb,amsfonts}%
\usepackage{amsthm}%
\usepackage{mathrsfs}%
\usepackage[title]{appendix}%
\usepackage[dvipsnames]{xcolor}
\usepackage[normalem]{ulem}

\usepackage{textcomp}%
\usepackage{manyfoot}%
\usepackage{booktabs}%
\usepackage{algorithm}%
\usepackage{algorithmicx}%
\usepackage{algpseudocode}%
\usepackage{listings}%
\usepackage{siunitx}
\usepackage{subcaption}
\usepackage{mathtools}
\usepackage{url}

\def\ket#1{\left|#1\right\rangle}
\def\ketSmall#1{\vert#1\rangle}

\def\ev#1{\left\langle#1\right\rangle}
\def\ketF#1{\left(#1\right)}

\newcommand{\abs}[1]{\left\lvert#1\right\rvert}

\newcommand{\ketbra}[1]{| #1\rangle \langle #1|}
\newcommand{\be}{\begin{equation}}
\newcommand{\ee}{\end{equation}}
\newcommand{\eea}{\end{eqnarray}}
\newcommand{\bea}{\begin{eqnarray}}
\newcommand{\va}[1]{\ensuremath{(\Delta#1)^2}}
\newcommand{\ex}[1]{\ensuremath{\langle{#1}\rangle}}
\renewcommand{\qed}{}

\newcommand{\REF}[1]{Ref.~\cite{#1}}
\newcommand{\eq}[1]{equation~\eqref{#1}}

\theoremstyle{thmstyleone}%
\theoremstyle{thmstyletwo}%

\theoremstyle{thmstylethree}%


\begin{document}

\title[Hong-Ou-Mandel interference of more than 10 indistinguishable atoms]{Hong-Ou-Mandel interference of more than 10 indistinguishable atoms}


\author*[1,2]{\fnm{Martin} \sur{Quensen}}\email{martin.quensen@dlr.de}
\author[1,2]{\fnm{Mareike} \sur{Hetzel}}\email{mareike.hetzel@dlr.de}
\author[1]{\fnm{Luis} \sur{Santos}}\email{luis.santos@itp.uni-hannover.de}
\author[3,4,5]{\fnm{Augusto} \sur{Smerzi}}\email{augusto.smerzi@ino.it}
\author[6,7,8,9,10]{\fnm{G\'eza} \sur{T\'oth}}\email{toth@alumni.nd.edu}
\author[3,4,5]{\fnm{Luca} \sur{Pezz\`e}}\email{luca.pezze@ino.cnr.it}
\author[1,2]{\fnm{Carsten} \sur{Klempt}}\email{carsten.klempt@dlr.de}

\affil[1]{\orgdiv{Institut f\"ur Quantenoptik}, \orgname{Leibniz Universit\"at Hannover}, \orgaddress{\street{Welfengarten 1}, \city{Hannover}, \postcode{30167}, \state{Lower Saxony}, \country{Germany}}}

\affil[2]{\orgdiv{Institut f\"ur Satellitengeod\"asie und Inertialsensorik (DLR-SI)}, \orgname{Deutsches Zentrum f\"ur Luft- und Raumfahrt e.V. (DLR)}, \orgaddress{\street{Callinstraße 30b}, \city{Hannover}, \postcode{30167}, \state{Lower Saxony}, \country{Germany}}}

\affil[3]{\orgdiv{Instituto Nazionale di Ottica}, \orgname{Consiglio Nazionale delle Ricerche (INO-CNR)}, \orgaddress{\street{Largo E. Fermi 6}, \city{Firenze}, \postcode{50125}, \state{Tuscany}, \country{Italy}}}

\affil[4]{\orgdiv{European Laboratory for Nonlinear Spectroscopy (LENS)},  \orgaddress{\street{Via Nello Carrara 1}, \city{Sesto Fiorentino}, \postcode{50019}, \state{Tuscany}, \country{Italy}}}

\affil[5]{\orgdiv{Quantum Science and Technologies in Arcetri (QSTAR)}, \orgaddress{\street{Largo E. Fermi 2}, \city{Firenze}, \postcode{50125}, \state{Tuscany}, \country{Italy}}}

\affil[6]{\orgdiv{Department of Theoretical Physics}, \orgname{University of the Basque Country UPV/EHU}, \orgaddress{\city{Bilbao}, \postcode{48080}, \state{Basque Country}, \country{Spain}}}

\affil[7]{\orgdiv{EHU Quantum Center}, \orgname{University of the Basque Country UPV/EHU}, \orgaddress{\street{Barrio de Sarriena}, \city{Leioa, Biscay}, \postcode{48940}, \state{Basque Country}, \country{Spain}}}

\affil[8]{\orgdiv{IKERBASQUE}, \orgname{Basque Foundation for Science}, \orgaddress{\street{Plaza Euskadi 5}, \city{Bilbao}, \postcode{48009}, \state{Basque Country}, \country{Spain}}}

\affil[9]{\orgdiv{Donostia International Physics Center (DIPC)},  \orgaddress{\street{Paseo Manuel de Lardizabal 4}, \city{San Sebasti\'an}, \postcode{20018}, \state{Basque Country}, \country{Spain}}}

\affil[10]{\orgdiv{HUN-REN Wigner Research Centre for Physics},  \orgaddress{\street{Konkoly-Thege Mikl\'os \'ut 29-33}, \city{Budapest}, \postcode{1121}, \state{Pest County}, \country{Hungary}}}


\abstract{When two indistinguishable bosons interfere at a beam splitter, they both exit through the same output port.
This foundational quantum-mechanical phenomenon, known as the Hong-Ou-Mandel (HOM) effect, has become a cornerstone in the field of quantum information.
It also extends to many indistinguishable particles, resulting in complex interference patterns.
However, despite of its fundamental and applied interest, the many-particle effect has only been observed in notoriously lossy photonic systems, but a realization with atomic systems has remained elusive until now.
Here, we demonstrate HOM interference with up to 12 indistinguishable neutral atoms in a system with negligible loss.
Our single-particle counting clearly reveals parity oscillations, a bunching envelope and genuine multi-partite entanglement, defining features of the multi-particle HOM effect.
Exploiting HOM interference, we demonstrate a Heisenberg scaling of metrological sensitivities achievable with the generated quantum states.
Our technique holds the potential to scale to significantly larger numbers, with wide-ranging applications from high-precision atom interferometry to multi-particle Bell tests.}




\maketitle

\noindent\makebox[\textwidth][c]{%
  \parbox{0.8\textwidth}{%
    \vspace{0.2em}%
    \small
    This version of the article has been accepted for publication, after peer review but is not the Version of Record and does not reflect post-acceptance improvements, or any corrections. The Version of Record is available online at: https://doi.org/10.1038/s41567-026-03302-7.
    \vspace{0.2em}%
  }%
}
\newpage
\clearpage
\newpage

\begin{figure}[ht]
	\centering
	\includegraphics{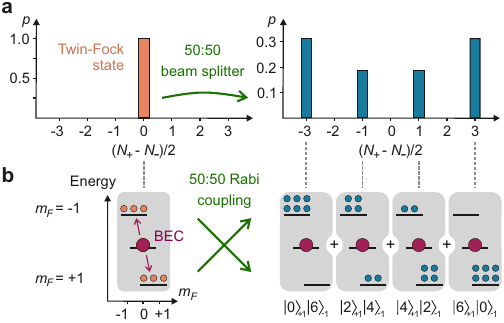}
    \caption{\label{fig1}\textbf{Multi-particle Hong-Ou-Mandel interference.} \textbf{a}, If $N_+=3$ and $N_-=3$ identical particles in a Twin-Fock state (left panel) interfere on a 50:50 beam splitter, the resulting state shows a characteristic distribution with only even particle numbers at the two output ports (right panel).
    The broad distribution of the number difference $(N_+ - N_-)/2$ has maximal contributions for all particles bunching at the same output port.
    \textbf{b} Spin-changing collisions within a Bose-Einstein condensate in the Zeeman level $\ketF{F,m_F}=\ketF{1,0}$ enable the creation of such Twin-Fock states in the two levels $\ketF{1,\pm 1}$ (left panel).
    A 50:50 Rabi coupling of these two levels is equivalent to a beam splitter and realises the same distribution of only even numbers.}
\end{figure}

Particle indistinguishably leads to striking quantum interference phenomena, as first demonstrated with a pair of photons in the landmark experiment by Hong, Ou and Mandel~\cite{hong_measurement_1987}.
The absence of simultaneous detection events at the two output ports of a balanced beam splitter serves as a central figure of merit for the generation of indistinguishable pairs.
The concept extends to bimodal states with a larger number of particles (Fig.~\ref{fig1}a) and their multi-particle interference.
If the same number of indistinguishable particles (Twin-Fock state) enter the two inputs of a 50:50 beam splitter, the output state exhibits a characteristic distribution of only even numbers with a bunching envelope~\cite{campos_quantum_1989}.
The HOM effect plays a key role in both quantum metrology~\cite{holland_interferometric_1993} and quantum information and has become a textbook example of interference phenomena which cannot be explained by a semiclassical theory~\cite{bouchard_two_2021}.
The Twin-Fock input states of the HOM interferometer are highly nonclassical quantum states that are particle-entangled according to the definition of Refs.~\cite{Sorensen2001,Morris2020}.
They offer enhanced sensitivity for interferometric applications~\cite{Campos2003}, surpassing the standard quantum limit and enabling Heisenberg scaling~\cite{lucapezze_quantum_2018}.


In optics, spontaneous parametric down-conversion (SPDC) is the prevalent method for creating indistinguishable photon pairs, integral to quantum cryptography~\cite{Gisin2002}, quantum computing \cite{OBrien2007}, quantum metrology~\cite{nagata_beating_2007,Polino2020}, and fundamental tests of quantum mechanics~\cite{Aspect2015}. 
Driving SPDC to higher photon numbers, multi-photon interference of four~\cite{ou_observation_1999,kiesel_experimental_2007}, six~\cite{prevedel_experimental_2009,wieczorek_experimental_2009,Xiang2013,Jin2016} and eight photons~\cite{Thekkadath2020} was observed, and entanglement-enhanced sensitivities were derived.
Although photon detectors have been developed with number-resolving capabilities extending to dozens of photons~\cite{Eaton2023}, photonic systems still encounter significant challenges, such as unwanted higher multiphoton components, partial distinguishability, and inevitable photon loss~\cite{Ollivier2021}; these imperfections affect the measured quantities~\cite{Ferreri2019} and compromise the comparison of model-free quantum mechanical predictions and experimental observations.
Furthermore, these effects strongly constrain the scaling to larger particle numbers~\cite{Tiedau2019}.

The generation of indistinguishable pairs has also been explored in other systems~\cite{Kaufman2018}.
HOM interference has been demonstrated with microwave photons~\cite{Lang2013}, ions~\cite{Toyoda2015}, Rydberg atoms~\cite{li_hong_2016}, and even between photons and atomic magnons~\cite{Wang2022}. 
Ultracold atomic systems enable HOM realizations through tunnel coupling in atomic tweezers~\cite{Kaufman2014} or in optical lattices~\cite{Lopes2015,Preiss2015a,Dussarrat2017}.
However, in these systems, multi-particle bosonic interference has been limited to multi-mode settings~\cite{Young2024,Islam2015}.
A direct extension of the original HOM interference from two particles in two modes to $N$ particles in two modes was not reported.
Below, we refer to this setting as multi-particle HOM interference.

A scaling and analysis of HOM interference to larger particle numbers is outstanding, which we realise with up to 12 indistinguishable neutral atoms, in a system with negligible loss and a single-particle resolving detection.
The HOM interference relies on the high-fidelity creation and detection of Twin-Fock states generated by spin-changing collisions in Bose-Einstein condensates (BECs). We analyze these states by evaluating parity observables, squeezing parameters as low as \SI{-15.4\pm0.1}{dB} and close-to-optimal multi-particle entanglement. We show that the created quantum state can be employed for entanglement-enhanced interferometry, and observe a Heisenberg scaling of the states' Fisher information for up to 12 atoms (corresponding to \SI{6.4\pm0.8}{dB} beyond the classical bound).

\section*{Pair creation in spinor Bose-Einstein condensates and single-atom counting}
Indistinguishable ultracold atoms in a gaseous Bose-Einstein condensate with a spin degree of freedom represent a paradigmatic many-particle quantum system near absolute zero temperature.
A BEC of atoms with a total spin $F=1$ shows a ground-state phase diagram characterised by three quantum phases~\cite{Ho1998,StamperKurn2013,yukikawaguchi_spinor_2012} with phase transitions driven by a pair creation process.
While this process generates states with equal particle numbers akin to SPDC~\cite{lucke_twin_2011,luo_deterministic_2017}, the correlations at the single-particle level were previously witnessed only indirectly.
Direct detection was hindered by detection noise, of typically a few atoms, stemming from technical noise in the recorded absorption or fluorescence signals.
With an improved fluorescence detection setup, it was possible to observe pair creation with a counting noise of $1.6$~atoms~\cite{Qu2020}, which is close to but not beyond the single-atom resolution threshold.
In our experiments, we reach a counting resolution of $0.2$~atoms, enabling direct observation of HOM interference in spinor BECs.

\section*{Experimental procedure}
We generate rubidium BECs of $250$ atoms in a crossed-beam optical dipole trap with waists of \SI{35}{\micro\meter} and  \SI{5}{\micro\meter}, respectively~\cite{Hetzel2024}.
The atoms are predominantly prepared in the Zeeman level $\ketF{F,m_F}=\ketF{1,0}$ by spin distillation cooling, and purified during a cleaning and compression sequence~(Methods).
When spin dynamics is initiated by tuning the quadratic Zeeman energy $q$ for a finite time $t=\SI{120}{ms}$, spin-changing collisions generate particles in pairs (Fig.~\ref{fig1}b), leading to a two-mode squeezed vacuum (TMSV) state~\cite{peise_satisfying_2015}
\begin{equation}
\ket{\xi}=\sum_{n=0}^{\infty} \frac{(-i \tanh{\xi})^n}{\cosh{\xi}} \ket{n}_{+1}\ket{n}_{-1},
\label{eq:squeezedvac}
\end{equation}
\noindent where $\xi=\Omega t$ is the squeezing parameter, which depends on the spin dynamics rate $\Omega= 2 \pi \times \SI{2.2}{Hz}$.
The notation $\ket{n}_{+1}\ket{m}_{-1}$ denotes a two-mode Fock state with $n$ atoms in the Zeeman level $\ketF{1,+1}$ and $m$ atoms in the Zeeman level $\ketF{1,-1}$, respectively.
The TMSV state constitutes a coherent superposition of Twin-Fock states with equal particle numbers in the two modes. 
Like in SPDC, the final atom number measurement will determine the state's total particle number $N$.
Because states with different total particle numbers do not interfere, we can treat the quantum state as a single Twin-Fock state prior to the HOM interference.
Interference is initiated by coupling the levels $\ketF{1,\pm 1}$ through a sequence of resonant microwave pulses, effectively achieving a 50:50 Rabi coupling~(Methods).
The coupling yields an output state according to
\begin{equation}
\ket{n}_{+1}\ket{n}_{-1} \rightarrow \sum_{k=0}^n c_k \ket{2k}_{+1}\ket{2n-2k}_{-1}, 
\label{eq:HollandBurnettState}
\end{equation}
\noindent where the coefficients $c_k$ follow a discrete arcsine distribution~\cite{campos_quantum_1989},  
\begin{equation}
\abs{c_k}^2 =  \binom{2 k}{k} \binom{2n-2k}{n-k} \left(\frac{1}{2}\right)^{2n}.
\label{eq:arcsine}
\end{equation}
It features a characteristic occupation of only even atom numbers and an enhanced probability of extremal states ($\ket{N}_{+1}\ket{0}_{-1}$ and $\ket{0}_{+1}\ket{N}_{-1}$), where most bosons occupy the same mode (bunching).
This Twin-Fock state after coupling is also called Holland-Burnett state~\cite{holland_interferometric_1993}.
Demonstrating the absence of odd occupation numbers in the final state constitutes a primary goal of our experiments and requires a single-atom-resolved counting of the occupation numbers in the levels $\ketF{1,\pm 1}$.

\begin{figure*}[ht!]
	\centering
	\includegraphics{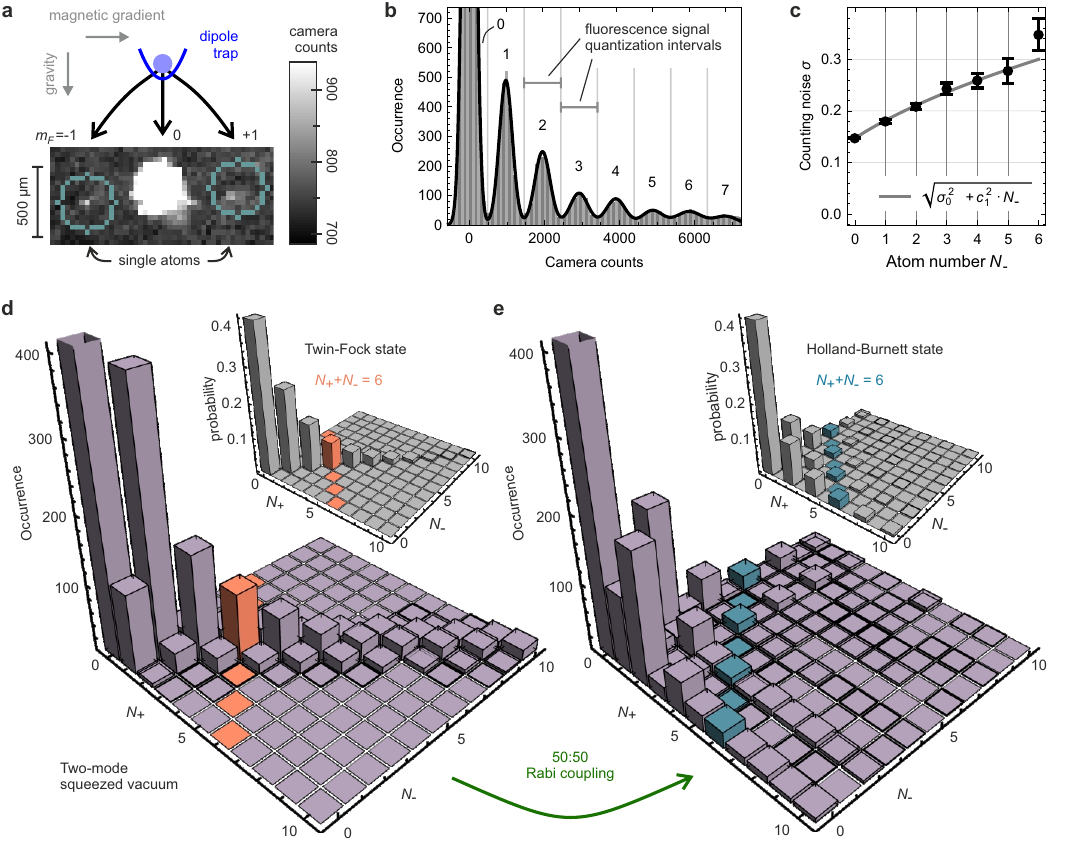}
    \caption{\label{fig2}\textbf{Accurate counting of atoms in a two-mode squeezed vacuum (TMSV) state.}
    \textbf{a}, A TMSV state is created in the levels $\ketF{1,\pm1}$ in an optical dipole trap.
    For counting, the atoms in these levels are spatially separated by a strong magnetic field gradient during free fall.
    They are illuminated with near-resonant laser light and imaged on a CCD camera.
    The displayed picture shows an exemplary fluorescence signal from two single atoms and the regions of interest used for signal summation (teal circles).
    \textbf{b}, The exemplary histogram of the measured camera signal of the $m_F=-1$ level (after subtraction of two systematic effects, see Methods) from 26712 repeated measurements features distinct peaks, demonstrating single-atom resolved counting.
    Gaussian fits provide a calibration of $\SI{975.8 \pm 1.6}{\mathrm{counts}/\mathrm{atom}}$ and, \textbf{c}, an atom counting noise $\sigma$ well below the single-atom level. The error bars represent the standard error of the fit parameters.
    The demonstrated single-atom resolved counting capability constitutes our main technical breakthrough that enables the observation of atomic HOM interference.
    \textbf{d}, The detection method is used to obtain the atom number distribution of the TMSV state.
    The preferential occupation of equal atom numbers $N_+=N_-$ (on the histogram's diagonal) is a characteristic of the ideal state (inset).
    \textbf{e}, After a 50:50 coupling of the two levels, the resulting histogram shows a characteristic checkerboard pattern with combinations of even numbers only (inset: ideal state).
    The subspace for a total number (highlighted in orange and blue for $N=6$) represents a Twin-Fock state (d) before and a Holland-Burnett state (e) after HOM coupling.
    Each histogram shows a data set of $3816$ repetitions.}
\end{figure*}

\section*{Accurate atom counting}
Single-atom-resolved counting is achieved using a fluorescence detection setup~\cite{hume_accurate_2013,huper_number-resolved_2021,Pur2023,Hetzel2023}, which is here operated in the light field configuration of an optical molasses~(Methods).
The setup consists of six intersecting laser beams with millimeter-sized diameters, red-detuned by ${2\pi \times \SI{6}{\mega\hertz}}$ from the Rb D2 cooling transition.
The fluorescence light from the atoms is imaged onto a CCD camera by a high-numerical-aperture lens system.
The detuning enables a long illumination time of \SI{4.2}{ms} due to continuous optical cooling, while the small beam diameters minimise stray light at our illumination intensities of $\sim\SI{3.6}{mW/cm^2}$. 
By spatially separating the atoms in the three Zeeman levels $\ketF{1, {-1}/0/{+1}}$ in a magnetic field gradient prior to illumination, Fig.~\ref{fig2}a, the detection becomes mode-resolved.
A single atom in $\ketF{1, \pm 1}$ causes a signal of approximately $900$ photoelectron counts on our CCD camera, whereas the recorded background noise is significantly lower, with a standard deviation of less than 0.17 atoms. 
The quantization of the camera signal to integer atom numbers, Fig.~\ref{fig2}b, demonstrates the desired single-atom-resolving counting capability.
Since fluorescence signals of dilute atomic gases do not exhibit any saturation effects~\cite{hume_accurate_2013}, which we checked for our system in a slightly modified, not mode-resolving configuration~\cite{Hetzel2023} for more than 100 atoms, we can extrapolate the quantization intervals linearly to larger atom numbers.
The counting noise, Fig.~\ref{fig2}c, is dominated by the background light's shot noise and camera noise for small atom numbers, and increases due to residual particle movement during the illumination~(Methods).
In the following, images are analyzed to assign integer atom numbers $N_-, N_0, N_+$ to the three Zeeman levels for each experimental realization.

\section*{Direct observation of many-particle Hong-Ou-Mandel interference}
First, the atom numbers are re\-cord\-ed for the generated two-mode squeezed vacuum state.
The histogram in Fig.~\ref{fig2}d shows how often a specific combination of atom numbers $N_-$, $N_+$ is observed during a measurement run with $3816$ identical repetitions.
It shows a dominant population of the Twin-Fock states with equal atom number on the diagonal.
There are minor off-diagonal contributions arising mainly from non-zero probabilities for miscounting.
Further contributions are unwanted state transfers during the spatial separation (main noise source for $N<3$), and losses prior detection.
A model of the noise contributions is provided in the Methods section.
The spin dynamics rate $\Omega$ undergoes small fluctuations that emerge from variations of the BEC atom number and the dipole trap configuration.
Consequently, the distribution of the total atom number in the Twin-Fock modes slightly deviates from the exponential distribution of \eq{eq:squeezedvac}.
The distribution in the subspace of a fixed, even total atom number $N = N_+ + N_-$ (highlighted for $N=6$) indicates, that the spin-changing collisions in our system serve as a high-fidelity source of atomic Twin-Fock states.
HOM interference, realised by a 50:50 Rabi coupling, fully reshapes the population distribution in accordance with \eq{eq:arcsine} (Fig.~\ref{fig2}e).
Detection events near the diagonal region are now least likely.

\begin{figure*}[ht!]
	\centering
	\includegraphics{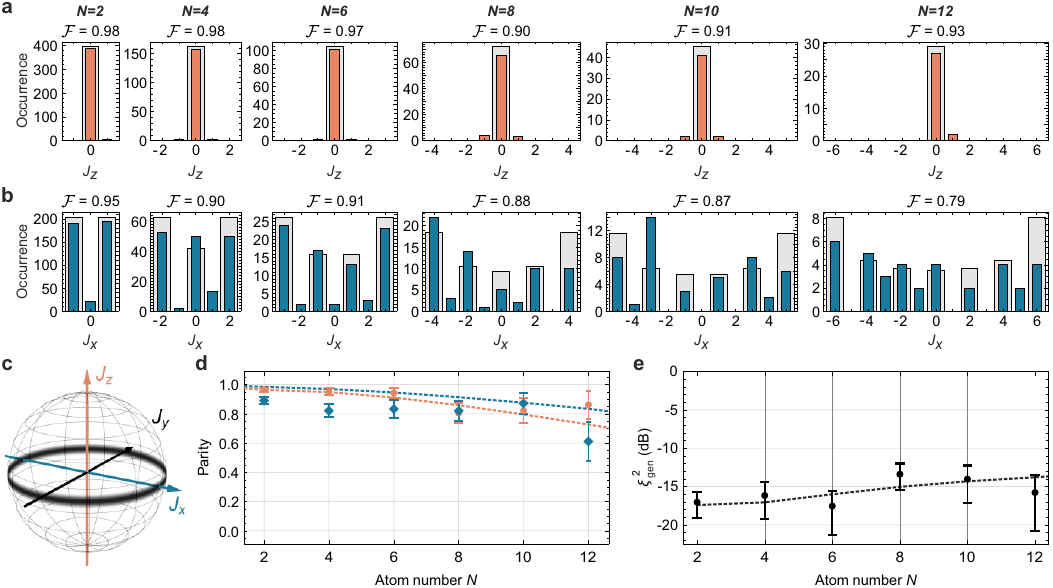}
	\caption{\label{fig3}\textbf{Demonstration of multi-particle Hong-Ou-Mandel interference.}
 The observed counting statistics before (\textbf{a}) and after (\textbf{b}) Hong-Ou-Mandel interference closely match those of the ideal states (gray bars), with fidelities $\mathcal{F}>0.87$ for up to $N=10$ atoms ($\mathcal{F}=0.79$ for $N=12$ atoms after coupling). 
 \textbf{c}, These two measurements constitute the observation of $J_z$ and $J_x$ of the initially prepared Twin-Fock state, respectively, illustrated by its Husimi distribution on the generalised Bloch sphere. 
 \textbf{d}, The observed parity signals remain near the maximum values of $\pm1$, and witness a clear suppression of odd occupation numbers after interference. This demonstrates the Hong-Ou-Mandel effect's main characteristic for up to $N=12$ particles. 
 \textbf{e}, The almost vanishing variance before HOM interference and the near maximum spread after coupling result in generalised squeezing parameters with an average value as low as \SI{-15.4\pm 1.0}{dB}.
 Error bars denote standard errors of the mean (d) and propagated statistical uncertainties (e). The dashed lines depict the predictions from our noise model (Methods).}
\end{figure*}

The non-classical features of the detected quantum states are highlighted in the distributions for selected total atom numbers $N$ (Fig.~\ref{fig3}).
The data is expressed in terms of the total angular momentum $\vec J=\sum_{n=1}^N {\vec j}^{(n)}$, defined as the sum of the spin-1/2 operators ${\vec j}^{(n)}$ of the $n$th atom in the two levels $\ketF{1, \pm 1}$. 
Note that a detection error by a single particle does not deteriorate these distributions, as the recorded total atom number $N$ becomes odd, and the event is discarded.
All of the following analyses are based exclusively on the directly observed population statistics, without incorporating any assumptions about underlying noise contributions.
Figure~3a shows the atom number difference $J_z = (N_{+} - N_{-})/2$ before the HOM interference.
The observed Twin-Fock states remain very close to the ideal state with $J_z=0$ even for $N=12$ atoms, with fidelities $\mathcal{F}>0.9$.
Figure~3b presents the number difference after HOM coupling, which is equivalent to a measurement of $J_x$ of the Twin-Fock input state.
$J_x$ and $J_y$ must be identical due to the state's symmetry and the absence of a phase relation between the atoms and the applied microwave pulses.
The fidelities of the states after HOM interference, calculated as $\mathcal{F}=\left(\sum_{k=0}^n \sqrt{p_k} \abs{c_k} \right)^2$ with $p_k$ the measured probabilities and $c_k$ the coefficients from \eq{eq:arcsine}, remain as high as $0.87$ for up to $10$ atoms ($\mathcal{F} = 0.79$ for $12$ atoms).
These states show a large suppression of odd-numbered mode occupations, Fig.~\ref{fig3}d, which can be quantified by the parity operator ${\Pi}_x$, where ${\Pi}_l=(-1)^{{N}/2-{J}_l}$ signals if the state exhibits even (${\Pi}_l=1)$ or odd (${\Pi}_l=-1)$ occupation numbers before ($l=z$) or after ($l=x$) HOM interference. 
We can directly assess the parity observable thanks to our number-resolved detection.
While parity oscillations have been observed with neutral atoms in optical lattices~\cite{Preiss2015a}, trapped ions~\cite{monz_14_2011} and Rydberg atoms~\cite{Omran2019}, our technique, using BECs, offers a particularly promising route to scaling to larger particle numbers.
We obtain values beyond $\pm 0.8$ for up to $10$ atoms and $0.6$ for $12$ atoms, revealing a key characteristic of the multi-particle HOM effect.
The very low variances $\Delta J_z^2 < 0.1$ of the Twin-Fock states and the large spread of ${J}_{x,y}$ after HOM interference lead to generalised squeezing parameters $\xi_\mathrm{gen}^2 = (N-1) \frac{\Delta J_z^2}{\ev{J_x^2+J_y^2}-N/2}$~\cite{lucke_detecting_2014,vitagliano_spin_2014} as low as \SI{-15.4\pm1}{dB} on average, Fig.~\ref{fig3}e.
While we do not directly observe a deterioration of $\xi_\mathrm{gen}^2$ for increasing atom numbers, our noise model (illustrated by the dashed lines) predicts a moderate decline.

We now further evaluate the quality of the performed HOM interference by extracting the multi-particle entanglement of the involved quantum state from the recorded data.

\begin{figure}[ht]
	\centering
	\includegraphics{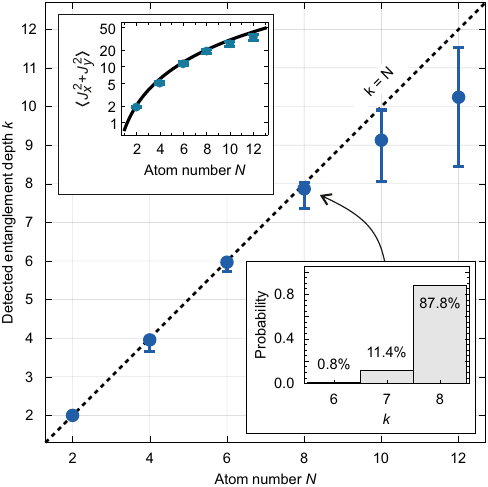}
 \caption{\label{fig4}{\bf Multi-particle entanglement.} The minimal number of entangled atoms $k$, that is compatible with the measured data according to \eq{eq:bipartsep_maintext}, demonstrates full genuine multi-particle entanglement for up to $N=8$ atoms.
 (top left inset) Measured (blue dots) and ideal (black solid line) width $\langle J_x^2+J_y^2\rangle$ of the occupation number difference after HOM interference, as presented in the distributions in Fig.~\ref{fig3}b.
 The error bars in the inset denote the standard error of the mean, while those in the main figure represent asymmetric standard deviations calculated via the Monte Carlo resampling approach~(Methods), illustrating the uncertainty in the computed distributions of the entanglement depth $k$, depicted here for $N=8$ (bottom right inset).}
\end{figure}

\section*{Multi-particle entanglement}
While the $J_z$ measurement, without HOM coupling, tests the equality of the particle numbers, the measurement of $J_{x,y}$, after HOM coupling, tests the coherence and the indistinguishability of the input state.
From these measurements, it is possible to quantify the multi-particle entanglement of the created Twin-Fock states.
This entanglement is quantified by the entanglement depth, defined as the number of particles in the largest nonseparable subset of the quantum state (Methods).
We evaluate a lower limit for the entanglement depth from the recorded data.
The inset of Fig.~\ref{fig4} shows the evaluation of $\langle J_x^2+J_y^2\rangle$ as a function of the total number of particles.
The large fluctuations of $\langle J_x^2+J_y^2\rangle$ are close to the optimal value of $\tfrac N 2 (\tfrac N 2 +1)$.
The slight deterioration for larger atom numbers is attributed to counting noise, loss, and fluctuations in the microwave transfer pulses.

From these measurements, we extract the entanglement depth according to a novel criterion~(Methods) that is based on the parity ${\Pi_z}$,
\begin{align}
&\ex{J_x^2+J_y^2}+\frac{k(N-k)}{2}|\ex{{\Pi}_z}|\le \frac N 2 \left(\frac N 2 +1\right).
\label{eq:bipartsep_maintext}
\end{align}
The parity directly probes $N$-body correlations within the ensemble.
If \eq{eq:bipartsep_maintext} is violated for some $k\ge N/2$, the entanglement depth is at least $(k+1).$
Figure~\ref{fig4} shows that the experimental data for up to $N=8$ atoms demonstrates full $N$-particle entanglement.
For larger numbers, the certified entanglement is not maximal, but no fewer than $10$ particles in the $N=12$ case with \SI{68}{\percent} confidence.
The Methods section contains a verification of the extracted entanglement depth by an alternative criterion~\cite{lucke_detecting_2014,vitagliano_entanglement_2017}.
It further provides an entanglement proof for the combined state for all $N\leq12$, including odd $N$, with a criterion for fluctuating particle numbers.
Here, we obtain an entanglement witness value of $-1.205(44),$  which is $27$ standard deviations beyond the classical bound of 0.
The presented single-atom resolved counting capabilities enable an entanglement certification and quantification that underlines the quality and quantum coherence of the generated mesoscopic Twin-Fock states.

\begin{figure*}[htp]
	\centering
	\includegraphics{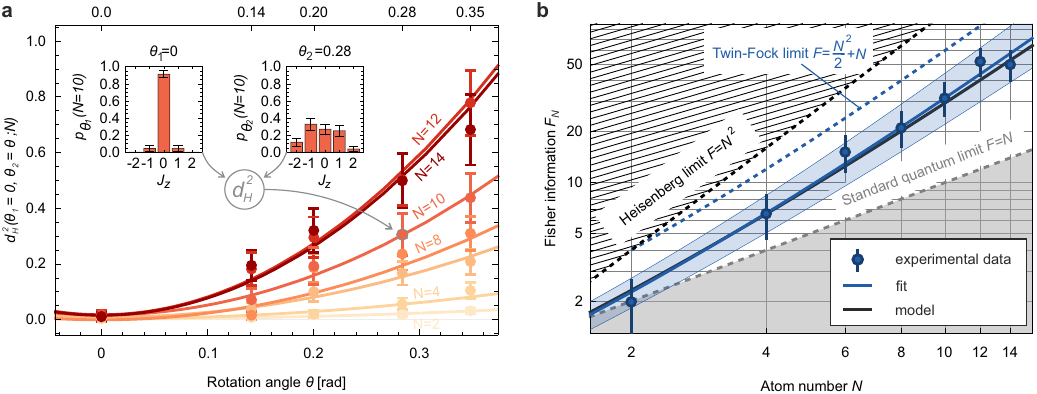}
	\caption{\label{fig5}{\bf Heisenberg scaling of the Fisher information with atom number.} \textbf{a}, Inset: For each atom number $N$ and rotation angle $\theta$, the probabilities for different $J_z$ values are estimated from the measured occurrences. The distinguishability of two probability distributions obtained at $\theta_1$ and $\theta_2$ is quantified by the squared Hellinger distance $d_H^2(\theta_1,\theta_2;N)$, \eq{eq:hellinger}. Main plot: The rate at which the Hellinger distance changes with varying phase differences $\theta_1 - \theta_2$ provides information about the state's metrological sensitivity. A quadratic fit $(F_N(\theta_1)/8) (\theta_1-\theta_2)^2$ to the calculated values of $d_H^2$ allows to extract the Fisher information $F_N(\theta_1)$. \textbf{b}, The weighted average of $F_N(\theta_1)$ over different $\theta_1$ is shown as a function of $N$. The obtained values for the Fisher information increase with atom number according to $r(\frac{N^s}{2}+N)$. The demonstrated scaling exponent of $s=\SI{1.95\pm0.09}{}$ is compatible with the theoretical prediction for ideal Twin-Fock states, $s=2$, with a factor $r=\SI{0.57\pm0.10}{}$ below the ideal values.
    The blue shaded area depicts the $68\%$ confidence region.
    Experimental data agree with a model including the combined effect of the dominant noise sources~(Methods), with no free fitting parameters.
    Error bars in (a) are obtained as standard deviations from a Monte Carlo resampling approach~(Methods), while those in (b) represent the averaged standard errors of the fit parameters $F_N(\theta_1)$.
 }
\end{figure*}

\section*{Metrological sensitivity}
The observed high-fidelity state creation, manipulation and detection can be exploited for interferometric applications.
Twin-Fock states can be utilized to achieve phase sensitivities close to the ultimate Heisenberg limit in systems with a single-atom resolved detection~\cite{Campos2003}. 
We quantify the achievable sensitivity of our $N$-particle Twin-Fock states through the evaluation of the Fisher information $F_N$.
The Fisher information is related to the Cram\'er-Rao bound $\Delta\theta_\text{CRB} = 1/\sqrt{F_N}$ for the estimation of a Rabi rotation of angle $\theta$~\cite{lucapezze_quantum_2018}.
By following a Hellinger method~\cite{strobel_fisher_2014},
$F_N$ is extracted from experimental data by measuring the squared statistical distance between probability distributions,  
\begin{equation}\label{eq:hellinger}
    d^2_H(\theta_1,\theta_2;N)=\sum_{J_z=-N/2}^{N/2} \frac{1}{2} \left(\sqrt{p_{\theta_1}(J_z;N)} - \sqrt{p_{\theta_2}(J_z;N)}\right)^2,
\end{equation}
where $p_\theta (J_z;N)$ is the relative population of the output value $J_z$ for a given rotation angle $\theta$.
A quadratic fit $d^2_H(\theta_1,\theta_2;N)=(F_N(\theta_1)/8) (\theta_1-\theta_2)^2$ provides an estimate of the Fisher information $F_N(\theta_1)$~(Methods).
In the noise-free case, Twin-Fock states allow reaching a Heisenberg scaling $F_N= N^2/2 + N$ with an improved robustness towards decoherence compared to highest-sensitivity NOON states.

We probe the metrological sensitivity by recording the relative population for $4$ rotation angles close to $\theta=0$ with another $4\times 3816$ experimental realizations, where the rotation angle quantifies the strength of the mode coupling.
The inset of Fig.~\ref{fig5}a shows exemplary histograms obtained for an $N=10$ Twin-Fock state before and after rotation by $\theta=\SI{0.28}{rad}$.
While the input state has negligible population of $J_z\neq 0$, these contributions increase with rotation angle: an effect that is captured by the Fisher information.
Figure~5a presents the squared Hellinger distance according to \eq{eq:hellinger} between the Twin-Fock state ($\theta_1=0$) and the rotated state for atom numbers ranging from $N=2$ to $N=14$, along with quadratic fits.
Figure~5b shows the mean Fisher information $\bar{F}_N$ obtained by averaging the fitting results $F_N(\theta_1)$ over the available angles $\theta_1$.
The obtained Fisher information significantly exceeds the classical limit $F_{N}=N$ for unentangled states: for $N=12$, we observe a Fisher information with $\SI{6.4\pm 0.8}{dB}$ enhancement.
Moreover, the Fisher information and the corresponding sensitivities increase with the same scaling as obtainable in an ideal measurement using Twin-Fock states.
To quantify this observation, we model the data using the function $\bar{F}_N=r\,\left(\tfrac{N^s}{2}+N\right)$, which is versatile enough to capture both the ideal Twin-Fock scenario (where $s=2$ and $r=1$) and the classical limit (where $s=1$ and $r=\tfrac{2}{3}$).
A fit yields $\bar{F}_N=0.57(10) \cdot (N^{1.95(9)}/2+N)$, which expresses a Heisenberg scaling of the interferometric sensitivity.
Entanglement-enhanced sensitivities for neutral atoms were demonstrated in thermal and condensed atomic ensembles~\cite{Pezze2018} with up to $20$~dB squeezing enhancement~\cite{Hosten2016} for $10^6$ atoms.
However, detection noise and other technical noise sources limited the achievable sensitivities to orders of magnitude above the Heisenberg limit. 
A Heisenberg scaling of phase sensitivity was so far only demonstrated with ions~\cite{monz_14_2011}.

\section*{Discussion}
The demonstrated HOM interference, including the generation and analysis of high-fidelity entangled many-particle states, opens a path towards quantum atom optics and atom interferometry with unprecedented fidelities, negligible losses and single-atom resolution.
Our setup and experimental capabilities allow for the further exploration of multi-partite entanglement in complex many-body systems, as those generated by crossing quantum phase transitions.
For instance, adiabatic passages enable the deterministic generation of Twin-Fock states~\cite{Luo2017,Anders2021} or even Schrödinger-cat-like states~\cite{Pezze2019}.
Furthermore, by realizing a spatial separation~\cite{Lange2018}, such states open the path to perform multi-particle Bell tests both with Twin-Fock~\cite{Laloee2009} and two-mode squeezed states~\cite{Ando2020,Lezama2023}.
The proposal of Ref.~\cite{Laloee2009} requires small losses and the ability to measure parity, both of which we demonstrated with the presented results.
Finally, the detection setup can be improved to detect up to 1000 atoms with single-atom resolution~\cite{hume_accurate_2013, Stroescu2015}.
The state generation can be improved to the same atom numbers by a reduction of the density and the corresponding three-body collisions.
Thus, our method promises to scale up the Heisenberg-limited sensitivities into an atom number regime, where the interferometric resolution becomes competitive with state-of-the-art unentangled sources, enabling a future generation of high-precision atom interferometers.

\newpage
\null\newpage
\renewcommand\figurename{Extended Data Fig.}
\renewcommand\tablename{Extended Data Table}
\setcounter{figure}{0}

\section*{Methods}\label{sec11}
\subsection*{Fluorescence detection}
For detection, the atomic ensemble is released from the cODT, exposed to a magnetic field gradient pulse during free fall, and then illuminated by six laser beams in optical molasses configuration, see Extended Data Fig.~\ref{figX1}a.
The magnetic field gradient is generated by a single coil, aligned with the homogeneous magnetic quantization field used for the coherent state manipulation.
\SI{0.5}{ms} after release, a capacitor is discharged through the coil over a period of \SI{6}{ms}, resulting in a peak current of \SI{430}{A} and a gradient of about \SI{40}{G/cm} at the position of the atoms.
A spatial separation of \SI{470}{\micro\metre} is reached for adjacent modes.
At the end of the pulse, \SI{6.5}{ms} into free-fall, both magnetic fields are turned off.
After another \SI{3.5}{ms}, during which the fields settle down, the laser beams of the optical molasses are turned on.

The molasses consists of six millimeter-sized, circularly polarised beams, detuned by one natural line width $\Gamma$ to the $F = 2 \rightarrow F' = 3$ cooling transition of the $^{87}$Rb D$_2$ line $5^2S_{1/2} \rightarrow 5^2P_{3/2}$.
The optical setup is the same as described in Ref.~\cite{Pur2023}, but with reduced beam diameters of the four horizontal beams from \SI{3.1}{mm} to \SI{1.1}{mm}.
This adaptation reduces the amount of stray light present during the illumination process, which is a main contribution to the counting noise. 
Beam intensities close to the saturation intensity for isotropically polarized light, $I_\mathrm{sat}=\SI{3.6}{mW/cm^2}$, result in an expected isotropic photon scattering rate of $R = \SI{1.04e7}{\mathrm{photons}/s}$.
Considering the photon collection efficiency of the detection's lens system of \SI{3.9}{\percent} and the properties of the employed CCD camera (Pixis 1024BR\_eXcelon\_WaterCool (1024x1024)), given by a quantum efficiency of $0.98$ primary electrons per incident photon and a set amplification gain of $0.92$ digital counts per electron, we expect $1530$~counts during the illumination time of \SI{4.2}{ms} per atom.
From the recorded data, we find values of $978$~counts/atom in $\ketF{1,-1}$, $1580$~counts/atom in $\ketF{1,0}$ and $830$~counts/atom in $\ketF{1,+1}$.
Since the molasses beams are aligned onto the position of the $\ketF{1,0}$ atoms after \SI{10}{ms} free-fall, these atoms experience the most intense and best intensity-balanced light field, and thus emit most fluorescence photons.
The Twin-Fock modes likely experience a slight intensity-imbalance of counter-propagating beams, as the horizontal displacement of \SI{470}{\micro\metre} is not negligible compared to the Gaussian beam waists of about \SI{550}{\micro\metre}.

\subsection*{Image evaluation}
The CCD camera array consists of $1024\times1024$ physical pixels.
Incident photons accumulate charge during the illumination time.
To reduce the level of electronic noise, arrays of $8\times 8$ are combined to a super-pixel ($\si{px}$ in the following) prior to readout, resulting in an image of size $\SI{128}{px}\times \SI{128}{px}$.
For the three modes $\ketF{1,-1}$, $\ketF{1,0}$ and $\ketF{1,+1}$, we add up the pixels' brightness values in constant areas (``masks'') around the respective bright spot to receive a single camera-count value for each image and mode, denoted $s_0$ and $s_\pm$ for $m_F=0$ and $m_F=\pm1$, respectively.
We found the best results for round areas with radii of \SI{5}{px}. Each mask includes $69$ (super-)pixels, with one (super-)pixel covering a physical area of $\SI{34.4}{\micro\metre}\times\SI{34.4}{\micro\metre}$ at the focus plane of the detection objective.
We take one image for each run of the experimental apparatus.
No background-image is taken.

A quantization of the signals, i.e.\ an accumulation of the obtained count numbers at evenly spaced values, is already visible in the raw data.
Importantly, the mean signals of the zero-atom peaks can easily be evaluated by a fit with a single Gaussian.
To receive the data shown in Fig.~\ref{fig2}b, we calculate and subtract the effects of two systematic effects (Extended Data Fig.~\ref{figX2}). 
Firstly, a very small but still relevant fraction of the $m_F=0$ atoms moves into the regions of the $m_F=\pm 1$ atoms during illumination and causes correlations of the zero-atom signals $s_{\pm}^{(N_\pm=0)}$ and the $\ketF{1,0}$ signal $s_0$, with correlation coefficients of \SI{1.48e-3}{} ($m_F=-1$) and \SI{1.76e-3}{} ($m_F=+1$).
Secondly, we find the zero-atom signals $s_{\pm}^{(N_\pm=0)}$, identified from 400 adjacent images, to drift slightly over the duration of the measurement, with the maximal and minimal observed signals being apart by 370 $\si{cts}$.

Finally, the occurrences of the count values are fitted with a sum of evenly spaced Gaussian functions,
\begin{equation}\label{eq:histogramFit}
G(s)=\sum_{n=0}^{n_\mathrm{max}+1} a_n \exp\left(\frac{(s-(n\, g + b))^2}{2 \sigma_n^2}\right),
\end{equation}
where $a_n$ are the peak heights, $\sigma_n$ the peak widths and $g$ the signal per atom. The position of the zero-atom peak $b$ is close to zero due to the applied signal drift correction. For quantization, each camera-count value per image and mode is now assigned the integer atom number $n$ of the closest peak, resulting in the quantization intervals depicted in Fig.~\ref{fig2}b and Extended Data Fig.~\ref{figX1}b, respectively. Note that this quantization technique extents to atom numbers much larger than $n_\mathrm{max}$, the number of the last peak that could be fitted.

For the fitting procedure, we weight the occurrences within each peak with the inverse of the total number of the peak's detection events.
To ensure convergence, the $n_\mathrm{max}+1$ peak is assigned a fixed width (which we predict from the widths of the previous peaks, iteratively).
The fits with \eq{eq:histogramFit} yield atomic fluorescence signals $g$ of \SI{832.5\pm3.4}{cts/atom} for the $m_F=+1$ mode and \SI{975.8\pm1.6}{cts/atom} for the $m_F=-1$ mode.
The counting noise is captured in the widths $\sigma_n$.
A noise model of the molasses detection predicts electronic camera noise and background light fluctuations (from shot noise and non-constant beam powers) to be the dominant contributions of the zero-atom signal noise, denoted $\sigma_0$.
The most relevant contribution that scales with atom number emerges from atoms leaving the detection volume during the illumination time due to the slowed, but not spatially restricted atom movement in the optical molasses.
This is captured by $c_1$~\cite{hume_accurate_2013}, and we get
\begin{equation}\label{eq:noiseFit}
    \sigma_n^2 = \sigma_0^2 + c_1^2 \cdot n.
\end{equation}

The fits with \eq{eq:noiseFit} to the obtained Gaussian widths of the peaks are presented in Fig.~\ref{fig2}c and Extended Data Fig.~\ref{figX1}c.
For the $m_F=-1$ mode we find $\sigma_0 = \SI{0.1466\pm0.0009}{\text{atoms}}$, incoherently increasing by $c_1 = \SI{0.107\pm0.003}{\text{atoms}/\sqrt{\text{atoms}}}$, while for $m_F=+1$ we find $\sigma_0 = \SI{0.168\pm0.004}{\text{atoms}}$ and $c_1 = \SI{0.164\pm0.015}{\text{atoms}/\sqrt{\text{atoms}}}$.
From this, we estimate detection fidelities, i.e.\ the chance of correctly counting the number of atoms, as the integral of the normalized Gaussian functions over the respective quantization intervals.
For up to $N=12$ atoms, we obtain fidelities of $>\SI{79}{\percent}$  for the $m_F=-1$ mode and $>\SI{60}{\percent}$ for the $m_F=+1$ mode.
The larger coefficient $c_1$ for $m_F=+1$ can be explained by a less optimal beam intensity balance at the position of these atoms after the spatial separation.

\subsection*{Coherent mode coupling and spin-changing collisions}
We employ a low-noise \SI{6.8}{GHz} microwave source~\cite{meyerhoppe_dynamical_2023} to drive Rabi oscillations between the $F=1$ and $F=2$ manifolds.
The transition frequencies are defined by an actively stabilised homogeneous magnetic field of $\SI{0.955}{G}$.
The microwave system was developed and tested for spin-squeezed samples of more than $10^4$ atoms, where it does not contribute relevant technical noise.
Rabi pulse lengths are typically in the order of \SI{100}{\micro\second}.
We implemented a spin-distillation scheme~\cite{a.couvert_quasi_2008} during the evaporative cooling, such that the BEC occupies the $\ketF{1,0}$ level while the side-modes $\ketF{1,\pm 1}$ are initially occupied by only a few atoms at most.
Prior to the spin-changing collisions, we remove these atoms by transferring them to $\ketF{2,\pm 1}$ and expose the ensemble to resonant cooling light.
We repeat this cleaning sequence three times.

We aim at small BEC atom numbers of around $250$ to avoid saturating the CCD camera during illumination and to reduce the effect of certain noise contributions, as discussed later.
To maintain a high spin dynamics rate of $\Omega = 2\pi\times\SI{2.2}{Hz}$, we increase the trap frequencies by raising the beam powers of the optical dipole trap from \SI{23}{mW} and $\sim\SI{320}{\micro W}$ to \SI{200}{mW} and \SI{2}{mW} after BEC creation.
The spin-dynamics rate scales as $\Omega \propto \overline{\omega}^{6/5} N_\mathrm{BEC}^{2/5}$ with the geometric mean $\overline{\omega}$ of the trap frequencies and the BEC atom number $N_\mathrm{BEC}$.
We apply a microwave dressing field on the clock transition to shift the Zeeman energy to resonance at $q = \hbar \Omega$~\cite{peise_satisfying_2015}.
After \SI{120}{ms}, we find a mean number of $7.5$~atoms in the levels $\ketF{1,\pm 1}$.
The distribution of the occupation numbers does not follow an exponential decay as predicted by \eq{eq:squeezedvac}.
We ascribe this discrepancy to a non-constant spin-dynamics rate due to fluctuations in atom numbers and trap frequencies.

For the coupling of the levels $\ketF{1, \pm 1}$, we apply a sequence of three microwave Rabi pulses (Extended Data Fig.~\ref{figX3}).
In between two $\pi$-pulses that transfer the atoms from $\ketF{1, -1}$ to $\ketF{2, 0}$ and back, we utilise a pulse of variable length on $\ketF{2, 0}\leftrightarrow \ketF{1, +1}$ to control the coupling ratio of the Twin-Fock modes $\ketF{1, \pm1}$.
Effective pulse lengths of \SI{7.66}{\micro\second}, \SI{10.8}{\micro\second}, \SI{15.4}{\micro\second}, \SI{18.8}{\micro\second} and \SI{85.0}{\micro\second} at a Rabi frequency of $\Omega_{\mathrm{R},\sigma^{-}} 
= 2\pi\times\SI{2.94}{kHz}$ result in small rotation angles between \SI{0.14}{rad} and \SI{0.35}{rad} and the HOM coupling corresponding to an angle of $\pi/2$.

Since the transition frequencies of $\ketF{1, 0}\leftrightarrow \ketF{2, \pm 1}$ only differ from those of $\ketF{1, \pm 1}\leftrightarrow \ketF{2, 0}$ by $\Delta = 2\pi\times\SI{2.66}{kHz}$, the pulse sequence also transfers some BEC atoms to $F=2$.
These atoms are removed by a cooling light exposure during free-fall, \SI{2.5}{ms} before the detection beams are turned on.
We observe that removing atoms this way can cause losses of $\ketF{1, \pm 1}$ atoms.
We thus keep the fraction of BEC atoms transferred to $F=2$ small by utilizing two different techniques.
For the coupling pulse, we choose a microwave antenna that couples $4.7$ times less to $\sigma^{+}$ than to $\sigma^{-}$ transitions, i.e.\ $\Omega_{\mathrm{R},\sigma^{+}}=\Omega_{\mathrm{R},\sigma^{-}}/4.7$, resulting in a maximally transferred fraction of $\Omega_{\mathrm{R},\sigma^{+}}^2/(\Omega_{\mathrm{R},\sigma^{+}}^2+\Delta^2) \approx \SI{2}{\percent}$.
For the two $\pi$ pulses, we choose the relative phase such that all BEC atoms, that were transferred by the first pulse, are transferred back to $\ketF{1,0}$ by the second pulse.

\subsection*{Calculation of fidelities $\mathcal{F}$}
For two probability distributions $p(J_z)$ and $q(J_z)$, the fidelity is given as $\mathcal{F} = \left(\sum_{J_z} \sqrt{p(J_z) q(J_z)}\right)^2$.
For the numbers given in Fig.~\ref{fig3}, we compare the experimentally observed probabilities $p^\textrm{exp}(J_z;N)$ with the probability distributions $q(J_z)$ of the ideal quantum states, where $J_z=-N/2,-N/2+1,...,N/2$ represents all possible values for a given even atom number $N$.
The experimental probabilities $p^\textrm{exp}(J_z;N)$ are obtained from the number of occurrences of the value $J_z$ during the full measurement.
The ideal states' probability distributions are given by $q(J_z) = \delta_{0,J_z}$ for the Twin-Fock states and by \eq{eq:arcsine} for the states after HOM interference.
Expressed in terms of $J_z$ and $N$, it reads
\begin{equation*}
    q(J_z) = \left\{\begin{array}{ll} \binom{N/2+J_z}{(N/2+J_z)/2} \binom{N/2-J_z}{(N/2-J_z)/2}\left(\frac{1}{2}\right)^{N}, & J_z+N/2\text{ even} \\
         0, & J_z+N/2\text{ odd}\end{array}\right..
\end{equation*}

\subsection*{Parity}
The parity operator for a single mode assigns a value of $+1$ to even occupation numbers and $-1$ to odd occupation numbers~\cite{Gerry2010}.
Since all states can be written as a superposition of Fock states $\ket{n}$, this property fully defines the operator.
It can be written as ${\Pi}_\text{single mode} = (-1)^{\hat{n}}$, with $\hat{n} = \ketbra{n}$ the occupation number operator.

For our two-mode system, in principle two parity operators exist, $\Pi_+ = (-1)^{\hat{N}_+}$ and $\Pi_- = (-1)^{\hat{N}_-}$.
However, for an even total atom number $N=N_++N_-$, we note that $(-1)^{N} \equiv 1$, such that
\begin{equation*}
    {\Pi}_z \coloneqq (-1)^{N/2-{J}_z} = (-1)^{{N}_-} = (-1)^{{N}_+}
\end{equation*}
is well-defined and describes the occupation number parity for both of the measured modes.
Similarly, we can define parity operators for all spin components $J_l$, with $l=x,y,z$, as
\begin{equation*}
    {\Pi}_l \coloneqq (-1)^{N/2-{J}_l}.
\end{equation*}
$\Pi_{x,y}$ describe parity measurements after rotating the state by \SI{90}{\degree} on the generalised Bloch sphere, i.e.\ after HOM interference.
Its relation to single-particle operators is given by ${\Pi}_l = \sigma_l^{\otimes N}$, with $\sigma_l = 2 j_l$ twice the $l$-component of the single-particle spin-1/2 operator, which can be written as a Pauli matrix and has eigenvalues of $\pm1$.

\subsection*{Probabilistic noise model of the measurements}
We deliberately avoid using assumptions about the underlying noise sources in the analysis of the recorded population data.
In the absence of any noise correction technique, the presented results directly describe the performance of our system.

For simulation purposes, to compare the results with theoretical expectations, and for consistency checks, we have developed a numerical model that describes the combined effect of various noise contributions on the population probabilities $p_\theta(J_z;N)$.
These probabilities for the possible outcomes $(J_z;N)$ are modeled as an array, $p^\textrm{model}_\theta(J_z;N)$, similar to those depicted in the insets of Fig.~\ref{fig2}d and Fig.~\ref{fig2}e.
To ensure that truncation effects are negligible, we choose the model to consider up to $N=20$ atoms per mode, which is a range significantly larger than that used for Fig.~\ref{fig3}, Fig.~\ref{fig4} and Fig.~\ref{fig5}.
We have checked that the results are stable with respect to array size changes.
The calculation of the probabilities follows these steps:
\begin{enumerate}
    \item Start with the probabilities of a superposition of Twin-Fock states, where the distribution of the total atom number $N$ follows the recorded average of all measurements.
    \item Change the probabilities within the subspaces of constant $N$ according to a rotation by the angle $\theta$. For example, $\theta=\SI{90}{\degree}$ results in Holland-Burnett states.
    \item Undesired transfers of atoms in the BEC reservoir into the detected modes $m_F=\pm1$ can occur after the mode interference, when magnetic fields get changed quickly before and during the strong magnetic field gradient pulse for spatial separation.\\
    These additional particles are modelled by a convolution of the probability array with Poisson distributions with parameters $a_{\pm}$.
    \item When those BEC atoms, that were transferred to $\ketF{F=2, m_F=\pm1}$ during the Rabi coupling sequence, Extended Data Fig.~\ref{figX3}, are removed by a short resonant light pulse after the magnetic field gradient, losses in $\ketF{F=1, m_f=\pm1}$ might occur due to collisions with the accelerated atoms. This effect is well visible when purposely removing a very large number of atoms.\\
    These losses are describes by convolutions with Binomial distributions with probabilities $1-l_{\pm}$.
    \item The calibration of the detection, i.e.\ the definition of the quantization intervals shown in Fig.~\ref{fig2}b and Extended Data Fig.~\ref{figX1}b, will have some error. Looking at measurement outcomes for Twin-Fock states with very high atom numbers $N$, we notice that the detection predicts slightly asymmetric numbers $N_-\approx 1.052 N_+$.\\
    This is modelled by applying chances of $\sqrt{1.052}$ to overpredict $N_-$ by one and underpredict $N_+$ by one. Please note that we cannot use this observation to refine the atom number assignment when analyzing the experimental data. This calibration method would assume, rather than demonstrate, the generation of Twin-Fock states.
    \item Finally, the finite detection resolution, i.e.\ counting noise, is considered by miscounting probabilities according to the overlap of the Gaussian peaks from \eq{eq:histogramFit} with adjacent quantization intervals.
\end{enumerate}
The model only has four free parameters, namely $a_{\pm}$ and $l_{\pm}$.

For each value of the rotation angle $\theta$ (\SI{0}{rad}, \SI{0.14}{rad}, \SI{0.20}{rad}, \SI{0.28}{rad}, and \SI{0.35}{rad}), we fit the model to the full array of measured experimental probabilities $p^\textrm{exp}_\theta(J_z;N)$ for $0\leq N_\pm \leq 20$, normalised such that
\begin{equation}
    \sum_{N=0}^{20} \sum_{J_z=-N/2}^{N/2} p^\textrm{exp}_\theta(J_z;N) = 1.
\end{equation}
Note that this analysis includes odd numbers of $N$, which are the primary effect of the noise contributions.
For the fitting, we minimise the Hellinger distance between $p^\textrm{model}_\theta(J_z;N)$ and $p^\textrm{exp}_\theta(J_z;N)$ by a differential evolution algorithm.

For each $\theta$, a total number of $3816$ experimental repetitions was carried out, each resulting in a pair $N_+$ and $N_-$ of detected atoms.
We obtain fit parameters of $a_+ = \SI{0.0551\pm0.0063}{}$, $a_-=\SI{0.0218\pm0.0018}{}$, $l_+=\SI{0.042\pm0.022}{\percent}$ and $l_-=\SI{1.1\pm0.8}{\percent}$ as the mean values for the five angles $\theta$.
The uncertainties are the statistical standard deviation.
We attribute the increased chance for losses in $m_F=-1$ to the asymmetry of the coupling sequence, as a small fraction of atoms might not be transferred back to $\ketF{F=1, m_F=-1}$ by the second $\pi$-pulse due to magnetic field fluctuations.

The obtained parameters indicate that losses have only a minor impact on our system.
The primary noise sources are finite detection resolutions, as depicted in Fig.~\ref{fig2}c and Extended Data Fig.~\ref{figX1}c, as well as unintended incoherent transfers from $\ketF{F=1, m_F=1}$ to $\ketF{F=1, m_F=\pm1}$, as described by $a_\pm$.

For clarity, please note that the results of the noise model are not incorporated in the data analysis for Fig.~\ref{fig3} to Fig.~\ref{fig5}.

\subsection*{Extracting the Fisher information from the Hellinger distance}
From the recorded occurrences for $(J_z;N)$ for the five small rotation angles $\theta$, we estimate the Fisher information $F_N$ when using the prepared $N$-atom Twin-Fock states from the experimental probabilities $p^\textrm{exp}_\theta(J_z;N)$, here normalised such that $\sum_{J_z} p^\textrm{exp}_\theta(J_z;N) = 1$.
Measured occurrences for $N=2$ and $N=10$ are displayed in Extended Data Fig.~\ref{figX4}.
The data clearly shows, that the rate, at which the measured experimental probabilities change with the rotation angle $\theta$, is much larger for the $N=10$ atom state.

To include an estimation of uncertainties in the analysis, we perform a Monte-Carlo resampling of the measurement occurrences that we use to calculate the experimental probabilities $p^\textrm{exp}_{\theta_{1,2}}(J_z;N)$.
We assume multinomial distributions with event probabilities given by $p^\textrm{exp}_{\theta_{1,2}}(J_z;N)$ for the $N+1$ possible outcomes of $J_z$.
Following the Monte-Carlo method, we compute the Hellinger distance of the recomputed distributions according to \eq{eq:hellinger}, and obtain the final value $d_{H,\mathrm{fit}}^2(\theta_1, \theta_2;N)$ as the mean and its uncertainty as the standard deviation.

For each possible choice of reference angle $\theta_1$, we fit the obtained values quadratically with
\begin{equation}
    d_{H,\mathrm{fit}}^2(\theta_1, \theta_2;N) = \frac{F_N(\theta_1)}{8} (\theta_1-\theta_2)^2 + b,
\end{equation}
where $F_N(\theta_1)$ and $b$ are free fit parameters.

Finally, we compute the weighted average of the Fisher information obtained for different $\theta_1$ as
\begin{align}
    \bar{F}_N   &= \sum_{\theta_1} \frac{w_{\theta_1}}{\sum_\theta w_\theta} F_N(\theta_1)\text{, with}\\
    w_{\theta_1} & = \left(\frac{1}{\Delta F_N(\theta_1)/F_N(\theta_1)}\right)^2.
\end{align}
Here, the weights $w_{\theta_1}$ are computed from the relative uncertainty of $F_N(\theta_1)$. For the uncertainty of $\bar{F}_N$, we use the averaged uncertainties of the fit parameters $F_N(\theta_1)$ rather than the standard error of the mean, as the five values $F_N(\theta_1)$ are not statistically independent.

We note that the resampling method introduces a statistical bias.
Since $d_{H,\mathrm{fit}}^2(\theta_1, \theta_2;N)$ is a convex function, this bias is positive.
Comparing the obtained mean values to those directly calculated from the measured experimental probabilities, we see that the bias only depends on the available sample sizes, but not directly on $\theta$.
As these sample sizes are almost independent of $\theta$, the free fit parameter $b$ can account for the introduced bias.

We further note that the Hellinger method itself is also affected by a statistical bias, see Ref.~\cite{strobel_fisher_2014}, especially when sample sizes are small.
To quantitatively determine the effect, we employ the probabilistic noise estimation model to predict probability distributions for all rotation angles $\theta$.
From these probabilities, the Hellinger distance is calculated directly, without any form of random sampling or rounding to integer occupation numbers.
Thus, no statistical bias is expected.
Employing the same fitting techniques as for the measured data, the model arrives at a scaling of $\tfrac{N^{1.81}}{2}+N$.
Adding the 4th order Taylor expansion term $-(\tfrac{1}{256} F_N^2 - \tfrac{1}{192} F_N) \theta^4$ to the fit function $d_{H,\mathrm{fit}}^2$, as expected for Twin-Fock states, results in $\tfrac{N^{1.87}}{2}+N$.
We further note that the measurement point for $\theta=\SI{0.35}{rad}$ and $N=14$ atoms lies outside of the range in which $d_{H}^2(\theta_1, \theta_2;N)$ can be approximated by a Taylor expansion, as there is a non-differentiable point at $\theta=\SI{0.321}{rad}$.
Neglecting this data point in the fit of $F_N$ yields a scaling of $\tfrac{N^{2.01}}{2}+N$.
Thus, our measurements are compatible with the $N^2$ scaling of the ideal Twin-Fock state.

\subsection*{Entanglement witness based on parity}
We consider the detection of entanglement in systems of many indistinguishable particles, using the definition of particle-entanglement as described, e.g., in Ref.~\cite{Morris2020}. The first important results in this field was the spin-squeezing entanglement condition based on first and second moments of collective observables~\cite{Sorensen2001}. Many experiments followed measuring collective quantities~\cite{Gross2010,Riedel2010,lucke_detecting_2014,Luo2017,Xin2023}.

In this section, we present entanglement relations that are based on $N$-particle correlations, rather than first and second moments of collective observables.

We can use the following witness to detect entanglement. For separable states 
\be
|\ex{\Pi_x}|+|\ex{\Pi_y}|+|\ex{\Pi_z}|\le 1\label{eq:Pixyz}
\ee
holds, which can be proved following ideas similar to those of Ref.~\cite{toth_detecting_2005}. For a product state of the type 
\be
\ketSmall{\Psi^{(1)}}\otimes \ketSmall{\Psi^{(2)}}\otimes ... \otimes\ketSmall{\Psi^{(N)}}
\ee
the left-hand side of \eq{eq:Pixyz} can be bounded from above as 
\be
\sum_{l=x,y,z}\left|\prod_{n=1}^N 
\ex{\sigma_l^{(n)}}\right|\le\sum_{l=x,y,z}|\ex{\sigma_l^{(1)}}\ex{\sigma_l^{(2)}}|\le 1
\ee
where in the first inequality we used that $|\ex{\sigma_l^{(n)}}|\le1,$ and in the second inequality we used the Cauchy-Schwarz inequality and the fact that the length of the Bloch vector is at most one for a qubit. Separable states are mixtures of product states, hence the inequality in \eq{eq:Pixyz} is also valid for separable states.

For the ideal Dicke state, for even $N,$ the left-hand side is three. It is a condition based on $N$-body correlations, unlike previous methods that were based on two-body correlations. Here, $\Pi_l$ is the parity operator from the main text and it equals $\sigma_l^{\otimes N}$  for $l=x,y,z.$

The  witness also detects the GHZ states as entangled. The singlet state given as $[(\ket{01}-\ket{10})/\sqrt{2}]^{\otimes N/2}$ has $\va{J_z}=0$ and $\ex{\sigma_x^{\otimes N}}=1,$ $\ex{\sigma_y^{\otimes N}}=1,$ if $N$ is divisible by $4.$ Thus, these operators cannot be used to detect genuine multipartite entanglement.

We summarise the results in Extended Data Table~\ref{tab:tab_xxxx}. We assume $\ex{\sigma_y^{\otimes N}}=\ex{\sigma_x^{\otimes N}}.$ The witness detects entanglement in all the cases. 

\begin{table*}[t!]
\centering
\begin{tabular}{
S[table-format = 2.1]
S[table-format = 2.4(2)]
S[table-format = 2.4(2)]
S[table-format = 3.4(2)]
S[table-format = 2.4(2)]
S[table-format = 2.5(2)]
}
\toprule
$N$ & $\ex{\sigma_x^{\otimes N}}$ & $|\ex{\sigma_z^{\otimes N}}|$ & $\ex{J_x^2+J_y^2}$ & $\mathcal{J}$ & $\va{J_z}$ \\ 
\midrule
 2  &  0.892(22) & 0.965(13) & 1.892\pm0.022 & 0.946\pm0.011 & 0.0176\pm0.0066\\
 4  &  0.821(44) & 0.951(25) & 5.08\pm0.29 & 0.85\pm0.05 & 0.025\pm0.012\\
 6  &  0.833(61) & 0.942(33) & 11.26\pm0.85 & 0.94\pm0.07 & 0.029\pm0.017\\
 8  &  0.821(70) & 0.806(70) & 19.0\pm1.6 & 0.95\pm0.08 & 0.098\pm0.036\\
 10 &  0.872(72) & 0.822(86) & 25.7\pm2.6 & 0.86\pm0.09 & 0.091\pm0.045\\
 12 &  0.61(13)   & 0.862(96) & 33.7\pm4.6 & 0.80\pm0.11 & 0.067\pm0.044\\
\bottomrule
\end{tabular}
\caption{\label{tab:tab_xxxx}Measurement results for various particle numbers. The uncertainties denote one standard deviation.}
\end{table*}

Next, we will derive an entanglement condition detecting genuine multipartite entanglement based on the parity operator $\Pi_z.$ As a first step we will derive a criterion detecting entanglement between two groups of the particles.

\subsection*{Entanglement condition using bipartite correlations}
In this section, we present a simple relation with expectation values of collective observables, as well as expectation values of $N$-particle correlations. We use these relations to obtain entanglement conditions based on bipartite correlations detecting entangement between two groups of particles.

{\bf Observation 1.} For $N$-qubit quantum states,
\be
\ex{J_x}^2/j^2+\ex{J_y}^2/j^2+\ex{\sigma_z^{\otimes N}}^2\le 1\label{eq:JxJyzzzz}
\ee
holds, where $j=N/2$ and 
\be
J_l=\frac 1 2 \sum_{n=1}^N \sigma_l^{(n)}
\ee
for $l=x,y,z.$

{\it Proof.} The ground state of the Hamiltonian
$
H=BJ_x+K\sigma_z^{\otimes N},
$
where $B$ and $K$ are constants, is of the form 
$
\ket{\Psi}=\alpha \ket{0}_x^{\otimes N}+\beta \ket{1}_x^{\otimes N},
$ 
which is a generalized Greenberger-Horne-Zeilinger (GHZ) state in the $x$-basis.
Then, the relevant expectation value of $J_x$ is
$
\ex{J_x}=\frac{N}{2} \ex{\sigma_x}_{\phi}
$ 
and the expectation value of the products of $\sigma_z$ matrices is
$
\ex{\sigma_z^{\otimes N}}=\ex{\sigma_z}_{\phi},
$
where we define the single-qubit state
$
\ket{\phi}=\alpha \ket{0}_x+\beta \ket{1}_x. 
$ 
 Since $\ex{\sigma_x}^2_{\phi}+\ex{\sigma_z}^2_{\phi}\le1,$ it follows that 
$
\ex{J_x}^2/j^2+\ex{\sigma_z^{\otimes N}}^2\le 1.
$ 
Then, assuming that the mean spin is not in the $x$-direction, but is in the $xy$-plane, we arrive at \eq{eq:JxJyzzzz}. $\qed$

{\bf Observation 2.} For bipartite separable states,
\be
\ex{J_x\otimes J_x}/(j_1j_2)+\ex{J_y\otimes J_y}/(j_1j_2)+|\ex{\sigma_z^{\otimes N_1}\otimes\sigma_z^{\otimes N_2}}|\le 1\label{eq:bipartite_corr}
\ee
holds, where $j_1=N_1/2$ and $j_2=N_2/2.$

{\it Proof.} We start from \eq{eq:JxJyzzzz} and use the Cauchy-Schwarz inequality, see e.~g.~Ref.~\cite{toth_detecting_2005}.$\qed$

\subsection*{Entanglement depth condition based on parity}
In this section, we obtain entanglement criteria detecting entanglement between two groups of the particles. We start from the relations based on bipartite correlations. Then, we obtain entanglement conditions that do not need bipartite correlations, but rather need the measurement of collective quantities.
Such quantities can be measured even in systems in which we cannot address the particles individually. Finally, we present a relation that detects the entanglement depth and can detect genuine multipartite entanglement.

{\bf Observation 3.}  The following expression is true for bipartite separable states
\begin{align}
&\sum_{l=x,y}\ex{(J_l^{(1)}+J_l^{(2)})^2}/(2j_1j_2)+|\ex{\sigma_z^{\otimes N}}|\le j(j+1)/(2j_1j_2),
\label{eq:bipartsep}
\end{align}
where $j_1=N_1/2, j_2=N_2/2,$ and $j=N/2.$

{\it Proof.}  We start from \eq{eq:bipartite_corr}. We add to both sides 
$
\sum_{l=x,y}\ex{(J_l^{(1)})^2}/(2j_1j_2)+\ex{(J_l^{(2)})^2}/(2j_1j_2).
$
Then follows the relation
 \begin{align}
&\sum_{l=x,y}\ex{(J_l^{(1)}+J_l^{(2)})^2}/(2j_1j_2)+|\ex{\sigma_z^{\otimes N}}|\\
&\le 1+\sum_{l=x,y}\ex{(J_l^{(1)})^2}/(2j_1j_2)+\ex{(J_l^{(2)})^2}/(2j_1j_2)\label{eq:bipartsep_Delta}
\end{align} 
Finally, we use the inequality
$
\ex{(J_x^{(n)})^2+(J_y^{(n)})^2}\le j_n(j_n+1).
$
$\qed$

Next, we will show how to use the criterion given in \eq{eq:bipartsep} for detecting genuine multipartite entanglement of the $N$-qubit system.

{\bf Observation 4.}  States violating the inequality given in \eq{eq:bipartsep} for $j_1=k/2$ and $j_2=(N-k)/2$ possess at least $(k+1)$-particle entanglement, where we assume that $k\ge N/2.$
Violation for $k=N-1$ means genuine multipartite entanglement. (cf.~\eq{eq:bipartsep_maintext})

{\it Proof.} The violation of \eq{eq:bipartsep} for $j_1=k/2$ means that the state cannot be written as a mixture of states of the form
\be
\ket{\Psi_1}\otimes\ket{\Psi_2},\label{eq:bisep}
\ee
where $\ket{\Psi_1}$ has $k$ qubits and $\ket{\Psi_2}$ has $(N-k)$ qubits. Note this is true for any groupings of the qubits into a group of $k$ and $(N-k)$ qubits. The states given in \eq{eq:bisep} are called biseparable states, since they are possibly multipartite entangled states that are separable with respect to a bipartition.

Without the loss of generality, let us consider the case $\ex{\sigma_z^{\otimes N}}\ge0$. Then, let us rewrite the inequality given in~\eq{eq:bipartsep} as 
\begin{align}
&\sum_{l=x,y}\ex{(J_l^{(1)}+J_l^{(2)})^2}+2j_1j_2\ex{\sigma_z^{\otimes N}}\le j(j+1).
\label{eq:bipartsep2}
\end{align}
The product $j_1j_2=j_1(j-j_1)$ is the largest for $j_1=j/2,$ and it is monotonously decreasing for a decreasing $j_1,$ and it is also monotonously decreasing 
if $j_1$ is increasing from $j_1=j/2.$ It is the smallest for $j_1=1/2, j_2=N/2-1/2,$ and for  $j_2=1/2, j_1=N/2-1/2.$ 
In general, the value of $j_1(j-j_1)$ is the same for $j_1$ as for $j-j_1.$
Hence, if the criterion \eq{eq:bipartsep2} is violated by a quantum state for $j_1\le j/2,$ then it is also violated for any $j_1'$ fulfilling $j_1\le j_1'\le j-j_1.$

Let us now consider the criterion in \eq{eq:bipartsep2} with $j_1=k/2,$
where we assumed $k\ge N/2$. Let us consider pure biseparable states of the form \eq{eq:bisep} such that $\ket{\Psi_1}$ has $N_{\Psi1}$ qubits and 
 $\ket{\Psi_2}$  has $N_{\Psi2}$ qubits, and $N_{\Psi1}\ge N_{\Psi2},$
 which also implies $N_{\Psi1}\ge N/2.$
Such a state can contain at most $N_{\Psi1}$-particle entanglement. Then, pure biseparable states of the form \eq{eq:bisep} with $N_{\Psi1}\le k$  cannot violate the criterion.
 Moreover, since \eq{eq:bipartsep2} is linear in expectation values, such a criterion cannot be violated even by states that are the mixtures of pure states of the type given in \eq{eq:bisep}, $\ket{\Psi_1}$ having $k$ or fewer qubits. Simple arguments then show that the criterion in \eq{eq:bipartsep2} cannot be violated by $k$-producible states, i.~e. states with at most $k$-particle entanglement. Thus, a state violating the criterion must possess at least $(k+1)$ particle entanglement or, equivalently, it must have at least an entanglement depth of $(k+1)$.

If a state violates the criterion given in \eq{eq:bipartsep2} for $j_1=1/2,$ then such a state cannot be a mixture of biseparable states of the form \eq{eq:bisep} with $\ket{\Psi_1}$ and $\ket{\Psi_2}$ being quantum states of one or more qubits. Thus, the quantum state must be genuine multipartite entangled. 

$\qed$ 

Observation 4 can be used to detect $(k+1)$-particle entanglement such that $k\ge N/2.$ We can also detect $k$-particle entanglement for $k<N/2$ as follows. The expectation values used for the entanglement criterion are shown in Extended Data Table~\ref{tab:tab_xxxx}. 
In the table, we also give the value of the parameter
\be
\mathcal{J}=\frac{\langle J_x^2+J_y^2\rangle}{N(N+2)/4},
\ee
which characterises how symmetric the state is. $\mathcal{J}=1$ corresponds to perfect bosonic symmetry. 

\subsection*{Entanglement depth condition based on collective measurements of spin components}
Let us use the definitions of the angular momentum components
\be
J_l=\frac1 2\sum_{n=1}^N \sigma_l^{(n)}
\ee
for $l=x,y,z.$
An entanglement condition has been defined in \REF{lucke_detecting_2014}. We use a somewhat stronger inequality in \REF{vitagliano_entanglement_2017},
due to which for states with at most $k$-particle entanglement the following inequality holds
\be
\va{J_z}\ge J_{\max} \mkern1mu{F_\frac{k}{2}}\mkern-7mu\left(\sqrt{\frac{\ex{J_x^2+J_y^2}-J_{\max}(\frac k 2+1)}{J_{\max}(J_{\max}-\frac k 2)}}\right),\label{eq:confvarJzNJP}
\ee
where the maximal spin length is defined as
$
J_{\max}=N/2,
$ 
$F_{j}(.)$ is defined in \REF{Sorensen2001a}.
If the above condition is violated, we have at least $(k+1)$-particle entanglement.

The criterion in \REF{lucke_detecting_2014} can be improved in a different way in certain cases. We can detect $(k+1)$-particle entanglement with a new condition given as
\be
\va{J_z}\ge J_{\max} \mkern1mu{F_\frac{k}{2}}\mkern-7mu\left(\frac{\sqrt{\ex{J_x^2+J_y^2}-X }}{J_{\max}}\right)\label{eq:cond}
\ee
Here we used the fact that the maximum of $\va{J_x}+\va{J_y}$ for pure states that are at most $k$-particle entangled is \cite{Hyllus2012a,Toth2012}
\be
X=\lfloor N/k\rfloor R(k)+R(r)
\ee
where we define
\be
R(n)=\begin{cases} n/2(n/2+1),  & \text{ even } n,\\
n/2(n/2+1)-1/4,  & \text{ odd } n.
\end{cases}
\ee
For an $n$-qubit state, $\va{J_x}+\va{J_y}\le R(n)$
holds.
Moreover, 
\be
r=N-\lfloor N/k\rfloor k.
\ee
We used both conditions on the experimental data, while for $k=1$ we used the condition for entanglement in \cite{Toth2007},
and chose the results of the method that gave a larger entanglement depth.
The results are shown in Extended Data Fig.~\ref{figX5}, the expectation values used for the criterion are given in Extended Data Table~\ref{tab:tab_xxxx}.

Note that for states that live in the $J_z=0$ subspace, $\ex{J_x}=\ex{J_y}=0,$ and for states with at most $k$-particle entanglement
\be
\ex{J_x^2+J_y^2}\le \lfloor N/k\rfloor R(k)+R(r)\label{Jx2Jy2}
\ee
holds. Any state that violates \eq{Jx2Jy2} is at least $(k+1)$-particle entangled.

\subsection*{Entanglement witness for an indefinite particle number state} 
For separable states with a given number of particles, we have \cite{Toth2007}
\be
(N-1)(\Delta J_z)^2-\left\langle J_x^2+J_y^2\right\rangle+\frac{N}{2}\ge0.\label{eq:critvarN_1}
\ee
Note that the left-hand side is identical to zero for $N=0$ and $N=1$.

Then, for the case of fluctuating particle numbers we use \cite{Hyllus2012}
\bea
(\Delta J_z)^2&-& \left\langle (N-1)^{-1} J_x^2\right\rangle
-\left\langle (N-1)^{-1}J_y^2\right\rangle\nonumber\\
&+&\frac{1}{2}\left\langle(N-1)^{-1}N\right\rangle\ge0,\label{eq:critvarN_1_fluctuating}
\eea
for which the left-hand side is $-1.205(44),$ more than $27$ standard deviations below zero. 
For the computation of the expectation values, we used the data for $N=2,3,4,...,12$ particles, which includes the odd particle numbers.

\subsection*{Calculation of uncertainties} 
Unless stated otherwise, the error bars throughout this article represent the standard errors.
For further details of our calculations of uncertainties, see Ref.~\cite{lucke_detecting_2014}. 

In Fig.~\ref{fig4} and Extended Data Fig.~\ref{figX5}, uncertainties are calculated using a Monte Carlo resampling approach.
This method is analogous to that used in our Fisher information analysis. 
For each atom number $N$, we resample the occurrences of the $J_z$ values (without HOM coupling) and $J_x$ values (after HOM coupling).
This resampling utilises multinomial distributions, where the sample sizes and probabilities are derived from the measured data, specifically $p^\textrm{exp}_\theta(J_z;N)$, which is normalised such that $\sum_{J_z=-N/2}^{N/2} p^\textrm{exp}_\theta(J_z;N) = 1$.
From each obtained sample (indexed by $i=0,1,...,10000$), we calculate value pairs $\{\ex{J_x^2+J_y^2}, \ex{{\Pi}_z}\}_i$ for \eq{eq:bipartsep_maintext} and $\{\ex{J_x^2+J_y^2}, \va{J_z}\}_i$ for \eq{eq:confvarJzNJP} and \eq{eq:cond}.
We then compute two samples $\{k_i\}$ of entanglement depth values: one based on parity (for Fig.~\ref{fig4}) and one based on the variance of $J_z$ (for Extended Data Fig.~\ref{figX5}).
The displayed values $k$ are the average of the $\{k_i\}$, and their uncertainties are displayed using upper and lower standard deviations, described below in \eq{eq:asymVariance}.
Extended Data Table \ref{tab:tab4} shows the minimally verified entanglement depths for confidence regions of \SI{68}{\percent} and \SI{95}{\percent}, i.e.\ the largest possible integer $k$ such that $k_i\geq k_\mathrm{min}$ still holds for at least \SI{68}{\percent} or \SI{95}{\percent} of the samples.

The novel entanglement depth criterion, \eq{eq:bipartsep_maintext}, is only applicable for $k\geq N/2$.
For \SI{1.31}{\percent} ($N=10$) and \SI{6.81}{\percent} ($N=12$) of the resampled value pairs, the criterion could not detect entanglement.
In these cases, we used the criterion given by \eq{eq:confvarJzNJP} and \eq{eq:cond} to detect entanglement.

\begin{table}[h]
\caption{\label{tab:tab4}Minimally verified entanglement depth for confidence regions of \SI{68}{\percent} and \SI{95}{\percent}.}
\begin{tabular}{@{\extracolsep\fill}ccccc}
\toprule%
& \multicolumn{2}{@{}c@{}}{$k_{\SI{68}{\percent}}$} & \multicolumn{2}{@{}c@{}}{$k_{\SI{95}{\percent}}$} \\\cmidrule{2-3}\cmidrule{4-5}%
$N$ & $\ex{\Pi_z}$-cond.\footnotemark[1] & $\Delta J_z^2$-cond.\footnotemark[2] & $\ex{\Pi_z}$-cond.\footnotemark[1] & $\Delta J_z^2$-cond.\footnotemark[2] \\
\midrule
2 &  2  & 2 &  2  & 2 \\ 
 4 &  4  & 4 &  4  & 4 \\ 
 6 &  6 & 6 &  6 & 6 \\ 
 8 &  8 & 7 &  7 & 6 \\ 
 10 & 9 & 8 & 7 & 7 \\ 
 12 & 10 & 9 & 7 & 7 \\ 
\botrule
\end{tabular}
\footnotetext[1]{According to \eq{eq:bipartsep_maintext}, with \eq{eq:confvarJzNJP} and \eq{eq:cond} only as fall-back options.}
\footnotetext[2]{According to \eq{eq:confvarJzNJP} and \eq{eq:cond}.}
\end{table}

For the asymmetric error bars employed in Fig.~\ref{fig4} and Extended Data Fig.~\ref{figX5}, we use the definitions for the upper and lower variance
\bea
(\Delta_+ x)^2&=&\frac2 M\sum_{n:x_n \ge\langle x\rangle} (x_n-\langle x\rangle)^2,\nonumber\\
(\Delta_- x)^2&=&\frac2 M\sum_{n:x_n <\langle x\rangle} (x_n-\langle x\rangle)^2,\label{eq:asymVariance}
\eea
where $x_n$ for $n=1,2,...,M$ are a set of values, and $\langle x\rangle$ is its average. With these definitions, for the variance
\be
(\Delta x)^2=\frac{(\Delta_+ x)^2+(\Delta_- x)^2}2
\ee
holds. 

\subsection*{Acknowledgements}
We thank I. Apellaniz and G. Vitagliano for discussions. We thank E. Rasel for the review of our manuscript. M.Q., M.H., L.S., and C.K. acknowledge financial support from the Deutsche Forschungsgemeinschaft (DFG, German Research Foundation), Project-ID 274200144, SFB 1227 DQ-mat within project B01, as well as Germany’s Excellence Strategy, EXC-2123 QuantumFrontiers, Project-ID 390837967. M.Q. also acknowledges support from the Hannover School for Nanotechnology (HSN). L.P. is supported by the QuantERA project SQUEIS (Squeezing enhanced inertial sensing), funded by the European Union’s Horizon Europe Program and the Agence Nationale de la Recherche (ANR-22-QUA2-0006). A.S. acknowledges funding from the Horizon Europe Program HORIZONCL4-2022-QUANTUM-02-SGA via the project 101113690 (PASQuanS2.1). G.T. acknowledges funding from the National Research, Development and Innovation Office of Hungary (NKFIH), Grant No. 2019-2.1.7-ERA-NET-2021-00036, the NKFIH within the Quantum Information National Laboratory of Hungary, and the European Union’s QuantERA projects MENTA and QuSiED. G.T. also acknowledges support from the Spanish MCIU, Grant No. PCI2022-132947, the Basque Government, Grant No. IT1470-22, and Grant No. PID2021-126273NB-I00 funded by MCIN/AEI and by “ERDF A way of making Europe.” Furthermore, G.T. acknowledges support from the “Frontline” Research Excellence Program of the NKFIH, Grant No. KKP133827, and Project No. TKP2021-NVA-04, implemented with support from the Ministry of Innovation and Technology of Hungary from the NKFIH.


\subsection*{Competing interests}
The authors declare that they have no competing interests.

\subsection*{Data and materials availability}
The raw fluorescence signal data, along with all other datasets used to generate the plots in this paper, are available in the LUIS data repository at \url{https://doi.org/10.25835/b77m3ufp} (Ref.~\cite{Quensen2025}).


\clearpage
\begin{appendices}

\begin{figure*}[tp]
	\centering
	\includegraphics{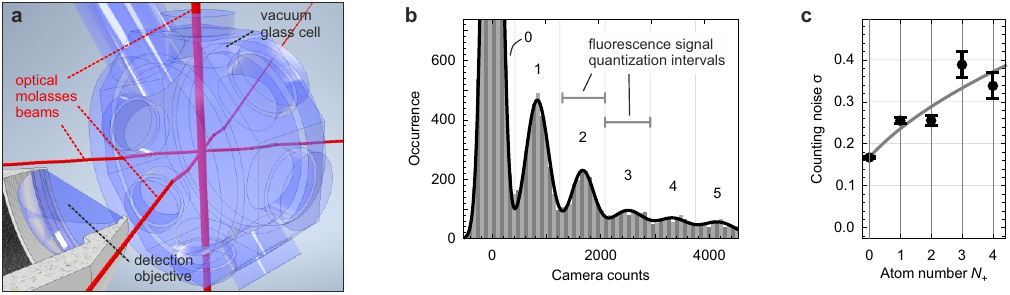}
	\caption{\label{figX1}{\bf Optical molasses configuration and detection resolution for $\mathbf{m_F=+1}$.} \textbf{a}, Six red-detuned beams in molasses configuration illuminate the atomic ensemble for detection. The fluorescence signal is captured by a detection objective and imaged onto a CCD camera. \textbf{b}, Histogram of the measured atomic signal of the $m_F=+1$ mode for all 26712 recorded measurements. The distinct peaks demonstrate the single-atom resolved counting capability. Gaussian fits of the peaks yield an atomic signal of \SI{832.5\pm3.4}{cts/atom}. \textbf{c}, The counting noise is quantified as the widths $\sigma_N$ of the Gaussian peaks. Error bars denote the standard errors of the obtained fit parameters. The slightly worse performance compared to the $m_F=-1$ mode is explained by a less optimal beam intensity balance at the position of the atoms after the spatial separation.}
\end{figure*}

\begin{figure*}[htp]
	\centering
	\includegraphics{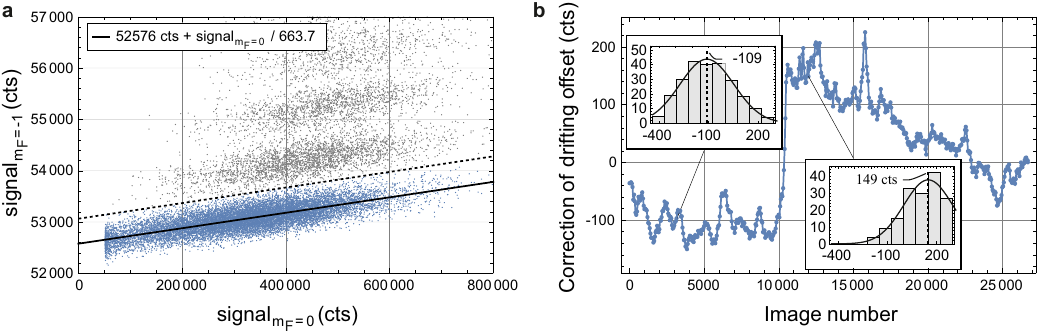}
	\caption{\label{figX2}{\bf Camera signal corrections.} \textbf{a}, Due to movement of $m_F=0$ atoms during the illumination, a small fraction of their fluorescence signal is present in the neighbouring masks. The correlation is captured with a linear fit (solid black line) to the signals of zero $m_F=-1$ atoms (blue points), and subsequently subtracted. \textbf{b}, A small drift of the background light intensity can be extracted from the recorded camera signals. For this, a single Gaussian function is fitted to the zero-atom signal peak of the histogram of 400 adjacent images. Subsequently, the peak position is subtracted from those images recorded at similar times.\\
    The data presented here is for the $m_F=-1$ mode. The data for the $m_F=+1$ mode looks similar.}
\end{figure*}

\begin{figure}[tp]
	\centering
	\includegraphics{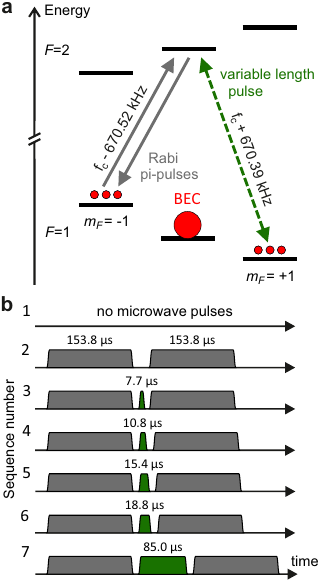}
	\caption{\label{figX3}{\bf Tunable mode coupling via microwave Rabi pulses.} (\textbf{A}) A sequence of three microwave pulses with frequencies around $f_c=\SI{6.835}{GHz}$ is employed for a variable Rabi coupling of the Twin-Fock modes $\ketF{1, \pm 1}$.
    The coupling strength is controlled by the duration of the second Rabi pulse. (\textbf{B}) For the measurements presented here, a total of seven different microwave sequences were repeated $3816$ times each.
    Sequence 1 and 2 both result in unmodified two-mode squeezed vacuum states. 
    Figures~2d and 3a show the data from sequence 2. 
    Sequences 3 to 6 constitute small rotations of the Twin-Fock states, employed for the results presented in Fig.~\ref{fig5}.
    Sequence 7 provides the data for the Hong-Ou-Mandel coupling, presented in Fig.~\ref{fig2}e and Fig.~\ref{fig3}b, and utilised for analysis results given in Fig.~\ref{fig3}d, \ref{fig3}e and Fig.~\ref{fig4}.}
\end{figure}

\begin{figure}[htp]
  \centering
    \includegraphics{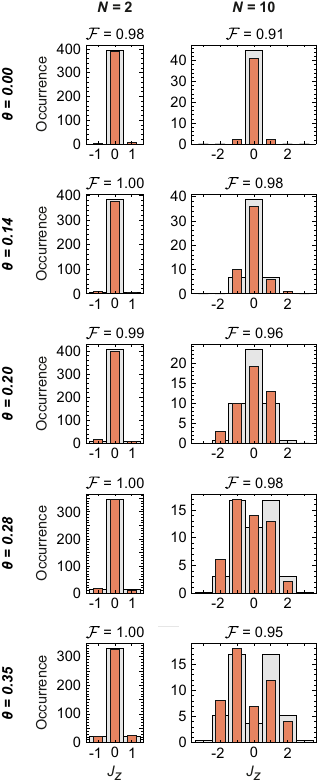}
    \caption{\label{figX4}{\bf Increased rate of change due to larger Fisher information.}
    The probabilities of different $J_z$ values change with the rotation angle $\theta$ (different rows).
    The measured occurrences (orange bars) very closely follow direct, model-free predictions (gray bars), indicated by fidelities $\mathcal{F}$ close to unity.
    The $J_z$ distribution for the $N=10$ state (right column) changes much faster than that of the $N=2$ state (left column), an effect that is quantified by the state's Fisher information $F_N$.
    }
\end{figure}

\begin{figure}[t]
	\centering
	\includegraphics{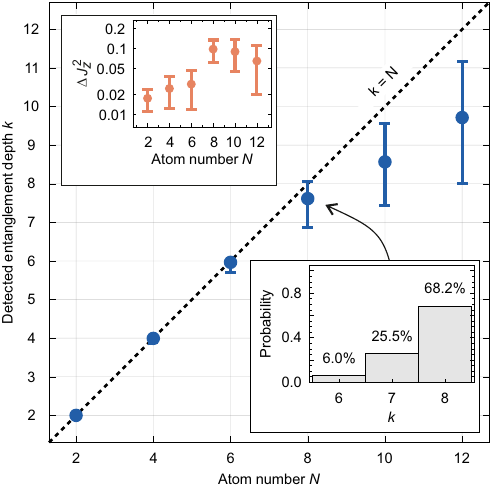}
 \caption{\label{figX5}{\bf Multi-particle entanglement.} The detected entanglement depth with the method based on variance $(\Delta J_z)^2$.
 The criteria are given in \eq{eq:confvarJzNJP} and \eq{eq:cond}. The error bars represent asymmetric standard deviations calculated via a Monte Carlo resampling approach, as described in section \textit{Error calculation}. 
 (Top left inset) The values of $(\Delta J_z)^2$ for various $N$. The error bars denote the standard error of the mean.
 (Bottom right inset) Each error bar represents a discrete distribution of $k$ values, shown here for $N=8$ as an example.}
\end{figure}

\end{appendices}
\end{document}